\documentclass[aps,showpacs,twocolumn,superscriptaddress]{revtex4}

\usepackage{graphicx}
\usepackage{dcolumn}

\begin{document}

\title{Comment on recent strangeness $\cal S$=$-$2 predictions}

\author{A.~Gal} 
\affiliation{Racah Institute of Physics, The Hebrew University, 
Jerusalem 91904, Israel} 

\begin{abstract} 
The Chiral Constituent Quark Model (CCQM)  interactions that bind 
the $H$ dibaryon and $_{\Lambda\Lambda}^{~~3}{\rm H}$ overbind 
$_{\Lambda\Lambda}^{~~6}{\rm He}$ by more than 4 MeV, thus outdating 
the CCQM in the $\cal S$=$-$2 sector. 
\end{abstract}

%few-body systems, hypernuclei, binding energies and masses, 
%nonrelativistic quark model

\maketitle

In a recent Letter, Garcilazo and Valcarce \cite{GV12} reported on 
a $\Lambda\Lambda N$--$\Xi NN$ coupled-channel three-body Faddeev 
calculation that binds $_{\Lambda\Lambda}^{~~3}{\rm n}$ and 
$_{\Lambda\Lambda}^{~~3}{\rm H}$ by about 0.5 MeV below the corresponding 
$\Lambda\Lambda N$ thresholds. This contrasts with {\it ab initio} 
$A\leq 6$ few-body coupled-channel calculations associating a loosely 
bound $_{\Lambda\Lambda}^{~~4}{\rm H}$ with the onset of $\Lambda\Lambda$ 
hypernuclear binding \cite{Nemura05}. 
Here I argue that the $\cal S$=$-$2 Chiral Constituent Quark Model (CCQM) 
interactions \cite{VGC10} that bind $_{\Lambda\Lambda}^{~~3}{\rm H}$ 
\cite{GV12}, as well as the unobserved $H$ dibaryon \cite{CV12}, overbind 
the uniquely identified NAGARA emulsion event of $_{\Lambda\Lambda}^{~~6}
{\rm He}$ \cite{NAGARA} by more than 4 MeV, casting doubts on the 
predictive power of the CCQM for $\cal S$=$-$2. 

\begin{table} [h] 
\caption{$\Delta B_{\Lambda\Lambda}({_{\Lambda\Lambda}^{~~6}{\rm He}})$=$
B_{\Lambda\Lambda}({_{\Lambda\Lambda}^{~~6}{\rm He}})$$-$$2B_{\Lambda}
({_{\Lambda}^{5}{\rm He}})$ (in MeV) from $\alpha\Lambda\Lambda$ calculations 
\cite{FG02,CAG97} with no $\Lambda\Lambda$--$\Xi N$ coupling, and scattering 
lengths $a_{\Lambda\Lambda}$ and effective ranges $r_{\Lambda\Lambda}$ (in fm) 
of the input $\Lambda\Lambda$ interaction $V_{\Lambda\Lambda}$. $\Delta 
B^{\rm exp}_{\Lambda\Lambda}({_{\Lambda\Lambda}^{~~6}{\rm He}})$=0.67$\pm$0.17 
MeV \cite{Nakazawa10}.} 
\label{tab:1} 
\begin{tabular}{lcccccccc} 
\hline \hline 
& \cite{FG02} & \cite{FG02} & \cite{FG02} & \cite{FG02} & \cite{FG02} & ~ & 
\cite{CAG97} & \cite{CAG97} \\ 
\hline 
$-a_{\Lambda\Lambda}$~ & 0.31 & 0.77 & 2.81 & 5.37 & 10.6 & ~ & 1.90 & 21.0 \\ 
$r_{\Lambda\Lambda}$~  & 3.12 & 2.92 & 2.95 & 2.40 & 2.23 & ~ & 3.33 & 2.54 \\ 
 
$\Delta B_{\Lambda\Lambda}$~ & 0.79 & 1.51 & 2.91 & 3.91 & 4.51 & ~ & 4.12 & 
8.29 \\ 
\hline\hline 
\end{tabular} 
\end{table} 

Listed in Table~\ref{tab:1} are $\Delta B_{\Lambda\Lambda}({_{\Lambda\Lambda}^
{~~6}{\rm He}})$ values obtained in two sets of $\alpha\Lambda\Lambda$ 
three-body calculations \cite{FG02,CAG97} which use identical $V_{\Lambda
\alpha}$; $V_{\Lambda\Lambda}$ from Ref.~\cite{CAG97} are softer than 
$V_{\Lambda\Lambda}$ from Ref.~\cite{FG02}. Within each set $\Delta B_{\Lambda
\Lambda}$ increases with increasing the strength of $V_{\Lambda\Lambda}$, 
as represented by the listed values of $-a_{\Lambda\Lambda}$. 
For $a^{\rm CCQM}_{\Lambda\Lambda}$=$-$3.3~fm, corresponding to the decoupled 
$V^{\rm CCQM}_{\Lambda\Lambda}$ \cite{Gar12}, interpolation within the first 
set \cite{FG02} suggests that $\Delta B^{\rm CCQM}_{\Lambda\Lambda}
({_{\Lambda\Lambda}^{~~6}{\rm He}})$=3.2$\pm$0.1~MeV, at variance with 
$\Delta B^{\rm exp}_{\Lambda\Lambda}({_{\Lambda\Lambda}^{~~6}{\rm He}})
$=0.67$\pm$0.17~MeV~\cite{Nakazawa10}. Interpolation within the second set 
\cite{CAG97} results in a value larger by at least 1 MeV. Since $V^{\rm CCQM}_
{\Lambda\Lambda}$ \cite{CV12} is softer than $V_{\Lambda\Lambda}$\cite{CAG97}, 
which is softer than $V_{\Lambda\Lambda}$\cite{FG02}, $\Delta B^{\rm CCQM}_
{\Lambda\Lambda}({_{\Lambda\Lambda}^{~~6}{\rm He}})$ should be even larger. 
Furthermore, the inclusion of the Pauli-suppressed $\Lambda\Lambda$--$\Xi N$ 
coupling increases $\Delta B_{\Lambda\Lambda}({_{\Lambda\Lambda}^{~~6}
{\rm He}})$ by another 0.2--0.5 MeV \cite{CAG97}, and by much more in the CCQM 
owing to its stronger coupling effects. Altogether I estimate conservatively 
$\Delta B^{\rm CCQM}_{\Lambda\Lambda}({_{\Lambda\Lambda}^{~~6}{\rm He}})>4.7
$$\pm$0.5~MeV, overbinding $_{\Lambda\Lambda}^{~~6}{\rm He}$ by more than 
4.0$\pm$0.5~MeV and thereby destroying the consistency among the bulk of 
$\Lambda\Lambda$ hypernuclear data \cite{GM11}. 

The CCQM $\Lambda\Lambda$--$\Xi N$ coupled-channel interactions used in 
Ref.~\cite{GV12} are not unambiguously constrained by the scarce, imprecise 
free-space scattering data \cite{Ahn06}. Figure~5 in Ref.~\cite{HM12} shows 
a variety of $\cal S$=$-$2 interactions satisfying such constraints. 
In particular, there are no $\Lambda\Lambda$ scattering data to constrain 
$a_{\Lambda\Lambda}$. Recent analysis of the $\Lambda\Lambda$ invariant mass 
from the in-medium reaction $^{12}$C($K^-,K^+\Lambda\Lambda X$) \cite{Yoon07} 
results in $a_{\Lambda\Lambda}$=$-$1.2$\pm$0.6~fm \cite{GHH12}, 
consistently with $a_{\Lambda\Lambda}$$\sim$$-$0.5~fm from 
$_{\Lambda\Lambda}^{~~6}{\rm He}$ \cite{FG02,VRP04}, in disagreement with 
$a^{\rm CCQM}_{\Lambda\Lambda}$=$-$3.3~fm. Furthermore, the very strong CCQM 
$\Lambda\Lambda$--$\Xi N$ coupling interaction which leads to a bound $H$ 
below the $\Lambda\Lambda$ threshold \cite{CV12} and is also responsible for 
binding $_{\Lambda\Lambda}^{~~3}{\rm H}$, is at odds with the latest HAL QCD 
lattice-simulation analysis which locates the $H$ dibaryon near the $\Xi N$ 
threshold \cite{HALQCD12}. For all these reasons, foremost for heftily 
overbinding $_{\Lambda\Lambda}^{~~6}{\rm He}$, the predictive power of the 
CCQM for $\cal S$=$-$2, including the prediction of a $_{\Lambda\Lambda}^{~~3}
{\rm H}$ bound state \cite{GV12}, is questionable.

\end{document}